\documentclass[a4paper,fleqn,usenatbib]{mnras}
\usepackage[utf8]{inputenc}
\usepackage[english]{babel}
\usepackage[T1]{fontenc}
\usepackage{ae,aecompl}
\usepackage{graphicx}
\usepackage{rotating} 
\usepackage{epsfig} 
\usepackage{amsmath}
\usepackage{amssymb}
\usepackage{euscript}
\usepackage{pdflscape}
\usepackage{xcolor}
\usepackage[normalem]{ulem}
\usepackage{url}
\title[RW~Aur\,B]{RW~Aur\,B: a modest UX~Ori type companion of the famous primary}

\author[A.~Dodin et al.]{
A.~Dodin,$^{1}$
S.~Lamzin,$^1$\thanks{E-mail: lamzin@sai.msu.ru}
P.~Petrov,$^2$
B.~Safonov,$^1$
M.~Takami,$^3$
and A.~Tatarnikov$^1$
\\
$^{1}$ Sternberg Astronomical Institute, Moscow M.V. Lomonosov State University,
Universitetskij pr., 13,  Moscow, 119992, Russia \\
$^{2}$ Crimean Astrophysical Observatory, Russian Academy of Sciences, 298409, Nauchny,
Crimea \\
$^{3}$ Institute of Astronomy and Astrophysics, Academia Sinica, 
11F of Astronomy-Mathematics Building, AS/NTU, 
No.1, Sec. 4, \\
Roosevelt Rd., Taipei 10617, Taiwan, R.O.C.%; hiro@asiaa.snica.edu.tw
}

\date{Accepted ~~~~~~~~~~~~~~~~~~~~~~~~~~~~~~~~. Received ~~~~~~~~~~~~~~~~~~~~~~~~~~~~~~; in original form }
\pubyear{2020}

\begin{document}
 \label{firstpage}
\pagerange{\pageref{firstpage}--\pageref{lastpage}}

\maketitle

\begin{abstract}
The secondary of the famous young binary RW~Aur is much less studied than the primary. To compensate this shortcoming, we present here the results of $UBVRIJHK$ photometric, $VRI$ polarimetric and optical spectral observations of RW~Aur\,B. The star demonstrates chaotic brightness variations in the optical band with irregular short $(\sim 1$ day) dimmings with an amplitude $\Delta V$ up to $1.3^{\rm m}.$ The dimmings are accompanied with an increase in the linear polarization (up to 3 per cent in the $I$ band), presumably due to the scattering of the stellar radiation by dust in the circumstellar disc that means that RW~Aur\,B can be classified as a UX~Ori type star. We concluded that the observed excess emission  at $\lambda \lesssim 0.45$~{\micron} and longward $\approx 2$~{\micron} as well as a variability of fluxes and profiles of \ion{H}{i}, \ion{He}{i} and \ion{Na}{i}~D  emission lines are due to the accretion process.  At the same time, emission components of \ion{Ca}{ii} lines indicate that RW~Aur\,B has a powerful chromosphere. Assuming the solar elemental abundances, we found the following parameters of the star: $T_{\rm eff} = 4100-4200$\,K, $A_{\rm V}=0.6 \pm 0.1$ (out of the dimming events), $L_* \approx 0.6$~$L_\odot,$ $R_* \approx 1.5$~$R_\odot,$ $M\approx 0.85$ M$_\odot,$ $\dot M_{\rm acc}<5\times 10^{-9}$ M$_\odot$ yr$^{-1}.$ Finally, we discuss possible reasons for the different levels of the accretion activity of RW~Aur binary components and present arguments in favour of the fact that they are gravitationally bound.
\end{abstract}

			%%%%%%%%%%%%%%%%%%%%%%%%%%%%%%%%%%%%%%%%%%%%%%%%%%%
%
\section{Introduction}
\label{sect:introduct}

RW~Aur is a young visual binary \citep{Joy-44} with the  present angular distance between the components of $\approx 1.5$\,arcsec \citep{Gaia-16b, Csepany-17}. The primary of the system RW~Aur\,A is a classical T Tauri star (CTTS), i.e. a low-mass young star, accreting matter from a protoplanetary disc \citep[][and references therein]{Petrov-01}.  The star has a bipolar jet (${\rm P.A.}=130\degr$), directed perpendicular to the major axis of the disc \citep{Hirth-94, Cabrit-06}.  \citet{Rodriguez-18} concluded that RW~Aur {\lq}has undergone multiple fly-by interactions{\rq} that means that it is a physical binary, such as the orbit of the companion RW~Aur\,B is highly elongated \citep{Bisikalo-12, Dai-15}.

Recently RW~Aur\,A has experienced two unusual dimming events.  The first one has occurred in 2010--11 ($\Delta t\sim 150^{\rm d},$ $\Delta V \sim 2$\,mag) and the second even deeper dimming started at 2014 summer and continues up to now \citep[][and references therein]{Rodriguez-18}. Comprehensive discussion of the behaviour of the star during these events one can find in the papers of \citet{Antipin-15}, \citet{Petrov-15}, \citet{Schneider-15}, \citet*{Shenavrin-15}, \citet{Bozhinova-16}, \citet{Facchini-16}, \citet{Takami-16}, \citet{Dodin-18b}, \citet{Garate-2018}, \citet{Gunther-18}, \citet{Koutoulaki+19}. While the nature of the dimmings is a matter of debates, there are no doubts that they as well as the jet have arisen due to the restructuring of inner regions of the circumprimary disc after the last close fly-by of the companion \citep{Berdnikov-17}.

Since RW~Aur\,B (the companion) is also a T~Tauri type star \citep{Joy-45} with protoplanetary disc \citep{Rodriguez-18}, it is important to know how the fly-by affected the behaviour of the star.  To answer this question, we present here results of our photometric, polarimetric and spectral observations of RW~Aur\,B.
%, carried out from 2014 to 2018.

The following is known about the star. According to \citet{Joy-45}: {\lq}The variable star RW~Aur has an 11.5-mag. visual companion $(d=1.2$\,arcsec; ${\rm P.A.}=254\degr$).{\rq} We found only three more estimates of visual brightness of the star in the literature prior to 2014: $V=13$ \citep[][Table X]{Herbig-1962}, $V>12.7$ \citep{GW67} and $V=13.63 \pm 0.05$ \citep{White-Ghez-01}. We also derived $V\approx 13.6$ from the spectrum shown in Figure 1 of \citet{Duchene-99}.  

\citet{Joy-45} reports that there are no strong bright lines in the blue spectrum of the star (1944, January, 4), {\lq}but in the region of $H\beta$ some emission is suspected.{\rq} \citet{Joy-Wilson-1949} estimated spectral type of the star as dM0e, and noted that single peaked H, K Ca\,II emission lines are present in its spectrum.  K3:e spectral type is specified in the \citet{Herbig-Rao-1972} and \citet{Herbig-Bell-1988} catalogs, but we believe that more reliable estimations of spectral type of RW~Aur\,B, based on the analysis of high resolution spectra, are K$6\pm1$ \citep{White-Hillenbrand-2004} and K6.5 \citep{Herzeg-Hillebrandt-2014}.

We found only two papers, where results of H$\alpha$ line equivalent width (EW) measurements in RW~Aur\,B spectra were presented: $W_{{\rm H}\alpha} = 42.7$ \AA{} \citep{Duchene-99}, $W_{{\rm H}\alpha} = 17$ \AA{} \citep{White-Hillenbrand-2004}.  Optical \citep{White-Hillenbrand-2004, Herzeg-Hillebrandt-2014} and near infrared \citep{Edwards-2006, Fischer-2008} spectra of the star are slightly veiled. According to \citet{Fischer-2008} profiles of hydrogen P$_\gamma$ and \ion{He}{i} $\lambda$1.083~\,{\micron} emission lines have redshifted absorption feature, indicating gas infall with velocity up to 400 km\,s${}^{-1}.$ Thus, it seems reasonable to agree with \citet{Duchene-99}, who concluded that the star can be {\lq}safely classified as CTTS{\rq}.

RW~Aur\,B has significant infrared excess longward 10\,{\micron} \citep{McCabe-06, Harris-2012, Andrews-2013}.  {\lq}A disturbed asymmetric peak{\rq} of emission in CO molecular lines and nearby 1.3 mm continuum were found around RW~Aur\,B by \citet{Cabrit-06} from {IRAM} observations.  \citet{Rodriguez-18} from {ALMA} observations concluded that the star has a Keplerian disc of radius $20-25$ au, and the discs of the A and B components of the binary have similar PAs, {\lq}but inclination of RW~Aur\,B disc axis to the line of sight $(i=67\degr \pm 1\fdg9)$ is $\approx 12\degr$ larger than that of the primary{\rq} (if their upper surfaces both face to the direction of the blueshifted jet of RW~Aur\,A). Later \citet{Manara-2019} found from {ALMA} observations in the 1.3 mm dust continuum the following values of the inclinations of the circumprimary and circumsecondary discs: $i_{\rm A}=55.0^{+0.5}_{-0.4}\degr$ and $i_{\rm B}=74.6^{+3.8}_{-8.2}\degr.$

The {\it Gaia} parallax for RW~Aur\,B ({\it Gaia} DR2 id 156430822114424576) is 
$6.12 \pm 0.07$ mas, which corresponds to the distance of $163 \pm 2$ pc
\citep{Gaia-16b}.

The rest of the paper is organized as follows. In Section 2 we describe our observations and present the results in Section 3.  Section 4 is devoted to interpretation of the results and in the last section we summarize our conclusions.
	            	%%%%%%%%%%%%%%%%%%%%%%%%%%%%%%%%%%%%%%%%%%%%%
%
\section{Observations}
 \label{sect:observation}

Resolved optical photometry of RW~Aur A and B was performed from 2014 November to 2020 March with the 2.5-m telescope of the Caucasian Mountain Observatory (CMO) of Sternberg Astronomical Institute of Lomonosov Moscow State University (SAI MSU) equipped with a mosaic CCD camera and a set of standard Bessel-Cousins $UBVRI$ filters.  One can find more detailed description of the equipment, observations and data reduction in our paper \citet{Dodin-18b}. Results of the resolved photometry for RW~Aur\,B are presented in Table\,\ref{tab:tab1}\footnote{rJD abbreviation in the first column of the Table means reduced Julian Date $\rm rJD=JD-2\,450\,000$ and will be used below as well.}.
%

			%%%%%%%%%%%%%%%%%%%%%%%%%
%
\begin{table}
\renewcommand{\tabcolsep}{0.095cm}
\caption{Optical photometry of RW~Aur\,B}
 \label{tab:tab1}
\begin{center}
\begin{tabular}{rcccccccccc} 
\hline
rJD     & $U$ & $\sigma_U$ & $B$ & $\sigma_B$ & $V$ & $\sigma_V$ 
& $R$ & $\sigma_R$ & $I$ & $\sigma_I$ \\
\hline
7077.20 &       &      & 14.81 & .02 & 13.53 & .02 & 12.57 & .05 & 11.47 & .02 \\
7376.42 &       &      &       &      &      &     & 12.26 & .05 &       &     \\
7986.53 & 14.89 & .05 & 14.21  & .02 & 13.00 & .02 & 12.07 & .05 & 11.07 & .02 \\
\hline
\end{tabular} \\
\end{center}
Tables \ref{tab:tab1}--\ref{tab:tab3} are available in their entirety
in a machine-readable form in the online journal.  A portion is shown in the
text for guidance regarding its form and content.

\end{table}
%
			%%%%%%%%%%%%%%%%%%%%%%%%%

Resolved near infrared (NIR) observations of RW~Aur\,A+B were carried out between 2015 December and 2020 March in the $JHK$ bands of the MKO photometric system at the 2.5-m telescope of CMO SAI MSU equipped with the infrared camera-spectrograph ASTRONIRCAM \citep{Nadjip-17}. Details of observations and data reduction were described by \citet{Dodin-18b}. Results of the observations are presented in Table\,\ref{tab:tab2}.

			%%%%%%%%%%%%%%%%%%%%%%%%%%%%%%
%
\begin{table}
\caption{NIR photometry of RW~Aur\,B} 
 \label{tab:tab2} 
 \begin{center}
\begin{tabular}{ccccccc}
\hline
rJD     & $J$ & $\sigma_J$ & $H$ & $\sigma_H$ & $K$ & $\sigma_K$ \\
\hline
7414.25 &  9.93 & .03 &  9.15 & .04 & 8.69 & .03 \\
7450.25 &  9.72 & .03 &  8.98 & .03 &      &      \\
7823.28 &  9.78 & .03 &       &     &      &      \\
\hline
\end{tabular}
\end{center}
\end{table}
%
			%%%%%%%%%%%%%%%%%%%%%%%%%%%%%%

Resolved polarimetric observations of RW~Aur\,A+B in the $VR_{\rm c}I_{\rm c}$ bands were carried out between 2015 October and 2020 April with the SPeckle Polarimeter (SPP) of the 2.5-m telescope of SAI MSU \citep{Safonov-17}.  Details of observations and data reduction  were described by \citet{Dodin-18b}.  Note that it was difficult enough to measure polarization of RW~Aur\,B: when the A component was in a bright state, the B component was much fainter, but during dimming events of the primary its flux was strongly polarised -- up to 30 per cent in the $I$ band \citep{Dodin-18b}. It appeared that the errors of measurements of RW~Aur\,B polarization are too large, when the degree of polarization $p$ of the star were less than 0.5 per cent.  Due to this reason we present in Table\,\ref{tab:tab3}, and will analyse below, only data with $p>0.5$ per cent in each band.

				%%%%%%%%%%%%%%%%%%%%%%%%%%%
%
\begin{table*}
\renewcommand{\tabcolsep}{0.15cm}
\caption{Optical polarimetry of RW~Aur\,B.}
 \label{tab:tab3} 
  \begin{center} 
\begin{tabular}{ccccccccccccccc} 
\hline 
rJD  &  $p_{\rm V}$ & $\sigma_{\rm p}$ & PA$_{\rm V}$ & $\sigma_{\rm PA}$
&  $p_{\rm R}$ & $\sigma_{\rm p}$ & PA$_{\rm R}$ & $\sigma_{\rm PA}$
&  $p_{\rm I}$ & $\sigma_{\rm p}$ & PA$_{\rm I}$ & $\sigma_{\rm PA}$
& $V$ & $\sigma_{\rm V}$ \\
& \% & \% & $\degr$ & $\degr$ & \% & \% & $\degr$ & $\degr$ & \% & \% & $\degr$ & $\degr$ & & \\ 
\hline 
7316.53 & 0.7 & 0.1  & 120  &  4& 1.0 &  0.1  &  122 &    3 &      &      &      &     & 12.79& 0.05 \\
8184.29 &     &      &      &   &     &       &      &      & 0.054& 0.02 &  50.5&  1.2& 13.47& 0.10 \\
8364.55 & 0.78& 0.20 &  69.0& 20& 0.41&  0.20 &  101 &   20 & 0.86 & 0.20 & 116.1& 20  &      &      \\
\hline 
\end{tabular}
\end{center}
Col.\,1: Date of observation;
Col.\,2--3: the polarization degree and its error in the $V$ band;  
Col.\,4--5: the polarization angle and its error in the $V$ band;
Col.\,6--9: the same as in col.\,2--5 but in the $R$ band;
Col.\,10--13: the same as in col.\,2--5 but in the $I$ band;
Col.\,14--15: $V$~magnitude and its error.  
\end{table*}
%
				%%%%%%%%%%%%%%%%%%%%%%%%%%%

We have three high resolution spectra of RW~Aur\,B at our disposal.  The first one was observed 1998 November, 6 ($\rm rJD=1123.568$) with the {SOFIN} spectrograph \citep*{Tuominen-1999} of the Nordic Optical Telescope.  The 3rd camera, which provides a spectral resolution of about 12 km\,s${}^{-1}$ with the entrance slit of 1.7 arcsec (R = 26 000), was used.  
%The star was observed within a set of RW~Aur\,A observations, so one can find more details in the paper of %\citet{Petrov-01}. 
 The spectrum of RW~Aur\,B was taken in a night of best seeing during a longer series 
of observations of RW~Aur\,A \citep{Petrov-01}.

Two additional spectra of RW~Aur\,B were observed 2016 January, 16 and 2018 January, 2 $(\rm rJD=7403.720$ and 8120.782, respectively) using the 3.6 m Canada-France-Hawaii Telescope equipped with the {ESPaDOnS} spectrograph, covering the wavelength range 3700--10\,500\,\AA. The instrument uses a $1.6$\,arcsec circular aperture. Seeing was 0.5 and 0.8 arcsec, respectively, at these observations. Details of the observations and data reduction are identical to that of RW~Aur\,A -- see \citet{Takami-16}. Additionally, we combined spectral orders and reduced the spectral resolution to $R=30\,000$ by convolution with the Gaussian profile. 

During the exposure of RW~Aur\,B spectrum,  RW~Aur\,A was always outside the entrance slit or the aperture of the spectrograph. Nevertheless, some scattered light from the primary contributed to the observed spectrum of the companion.  To decrease this contribution we took into account that emission lines in RW~Aur\,B's spectra are very few, weak and narrow (see Section~\ref{sect:spectroscopy}), while in the case of RW~Aur\,A they are much more numerous, stronger and broader. Therefore, we removed the spectrum of the primary (observed in the same night) weakened by the appropriate number of times from the spectrum of the companion until any traces of the broad emissions disappear. We suppose that in our spectra the contribution from RW~Aur\,A in the region between 4000 and 9000 \AA\, is less than $2-3$~per cent of the continuum intensity of RW~Aur\,B.  Note that at the moments of 1998 and 2018 observations RW~Aur\,B was 1.5-2 mag fainter (in the $V$ band) than RW~Aur\,A, but had about the same brightness as the A component during 2016 spectral observation -- compare our Fig.\,\ref{fig:fig1} and Figure.\,1 in \citet{Dodin-18b}.
        	    	%%%%%%%%%%%%%%%%%%%%%%%%%%%%%%%%%%%%%%%%%%%%%
%
\section{Results}
 \label{sect:results}
\subsection{Photometry and polarimetry}
 \label{sect:photometry}

\begin{figure}
 \begin{center}
\includegraphics[scale=0.45]{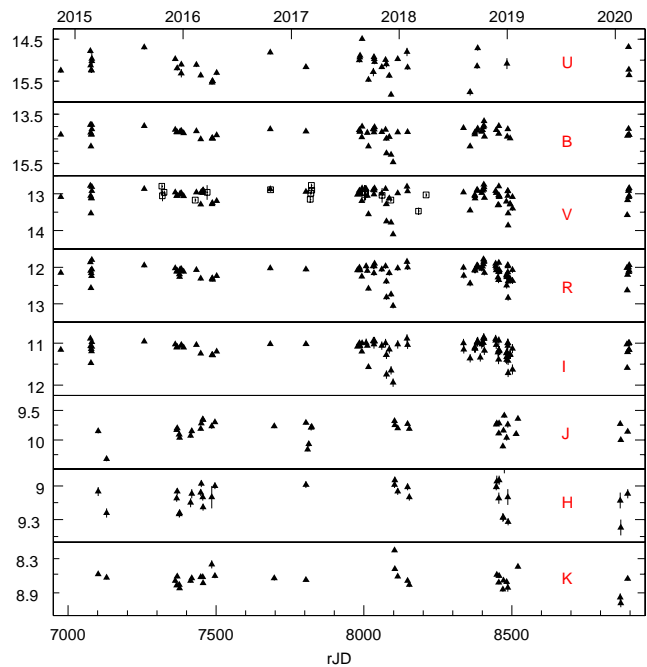}
 \end{center}
  \caption{The light curves of RW~Aur\,B in different spectral bands. The open squares in the $V$ band panel correspond to magnitudes derived from the SPP and AAVSO data (see text for details). The black triangles represent our resolved photometry, except the first two $JHK$ points $(\rm rJD\approx 7100),$ which were taken from \citet{Schneider-15}. The lower horizontal axis is for reduced Julian dates and the upper one is for calendar years, the ticks correspond to the beginning of each year.}
 \label{fig:fig1}
\end{figure}

The light curves of RW~Aur\,B in the $UBVRIJHK$ bands covering the period from 2014 September to 2020 March are presented in Fig.\,\ref{fig:fig1}.  The star is variable in all these bands, and the most interesting feature of its variability is irregular steep decreases of stellar brightness (dimmings).

				%%%%%%%%% Fig.lcBall %%%%%%%%%%%%
%
%
				%%%%%%%%%%%%%%%%%%%%%

\begin{figure}
 \begin{center}
\includegraphics[scale=0.46]{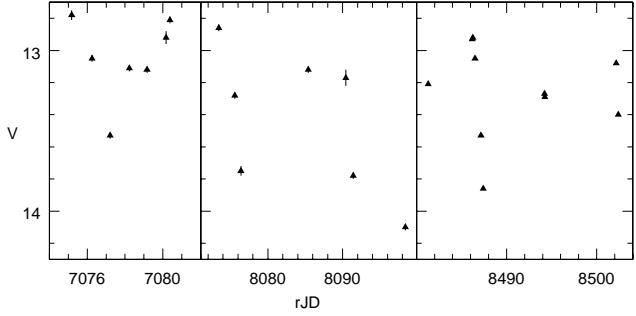}
 \end{center}
\caption{Portions of RW~Aur\,B's $V$ light curve in the vicinity of dimming events. 
}
\label{fig:dimmings}
\end{figure}

As can be seen from Fig.\,\ref{fig:dimmings}, the flux of the star in the dimming events in the $V$ band can decrease twice in less than a day and than come back to the initial level during 1-3 days.  Similar events were observed in other optical bands.  In general, the correlation coefficients $r$ between  magnitudes in the $V$ and $U,B,R,I,J,H $ bands are high -- 0.87, 0.99, 0.98, 0.97, 0.61, 0.76, respectively, but $r=0.02$ for the $K$ band.

			%%%%%%%%% Fig. dimming events in V band %%%%%%%%%%%%
%
%
			%%%%%%%%%%%%%%%%%%%%%
  
\begin{figure}
 \begin{center}
\includegraphics[scale=0.43]{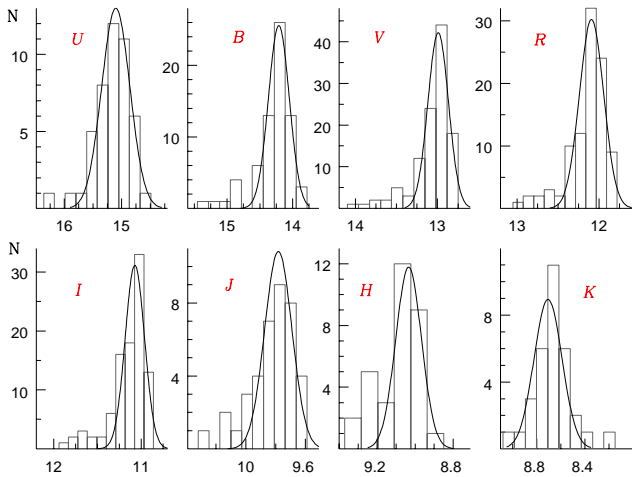}
 \end{center}
\caption{ The histogram of RW~Aur\,B brightness variations in the visual and NIR bands. The thin lines are for the Gaussian fits of the data.}
 \label{fig:hist}
\end{figure}

Histograms shown in Fig.\,\ref{fig:hist} demonstrate that the photometric behaviour of the star shortward and longward of $\lambda \approx 2$\,{\micron} is indeed different. In the $UBVRIJH$ bands the distribution of brightness can be described by a Gaussian curve centered at some bright state with a relatively long tail to the direction of low brightness, representing dimming events, but no such tail in the $K$ band is apparent.

			%%%%%%%%% Fig_Histogram  %%%%%%%%%%%%
%
%
			%%%%%%%%%%%%%%%%%%%%%
 
\begin{figure}
 \begin{center}
\includegraphics[scale=0.44]{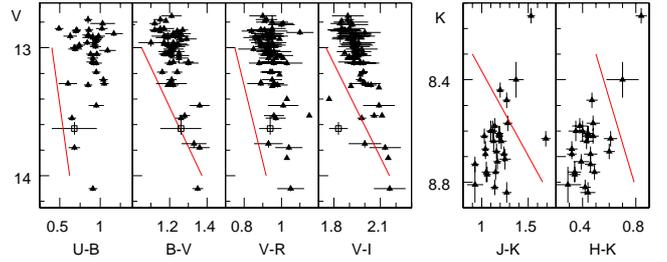}
 \end{center}
\caption{ The colour--magnitude diagrams for RW~Aur\,B in the visual (four left panels) and NIR (two right panels) bands. The black triangles are our observations, the open squares are observations of \citet{White-Ghez-01}. The red lines represent IS reddening.}
 \label{fig:colors}
\end{figure}

Colour-magnitude diagrams of RW~Aur\,B in the visual and NIR bands are shown in Fig.\,\ref{fig:colors}.  In the case of $V$ vs.  $B-V,$ $V-R$ and $V-I$ diagrams the weaker the star, the redder it is. But either the reddening is significantly less than the standard $R_{\rm V}=3.1$ interstellar (IS) one -- see the dashed line in the figure -- or these colour-magnitude tracks turn to the blue (the bluing effect) at $V>13.5$ due to scattered light. At the same time there is no correlation between the $U-B$ colour and brightness of the star: the correlation coefficient $r$ is $\approx -0.1,$ i.e. it looks like that the {\lq}reddening{\rq} is fully compensated by the {\lq}bluing{\rq}.

Only 16 out of 41 our observations of RW~Aur\,B in the NIR bands were carried out in the same nights with observations in the  visual bands.  Due to this reason we use the $K$ magnitudes instead of $V$ for NIR colour-magnitude diagrams shown in two right panels of Fig.\,\ref{fig:colors}.  One can see that in contrast to the visual band the brighter the star in the $K$ band the redder it is.  We will discuss this feature in Section~\ref{PandC-interp-2}.

			%%%%%%%%% Fig.colorsBopt %%%%%%%%%%%%
%
%\begin{figure*}
%\end{figure*}
%
			%%%%%%%%%%%%%%%%%%%%%

\begin{figure*}
 \begin{center}
\includegraphics[scale=0.75]{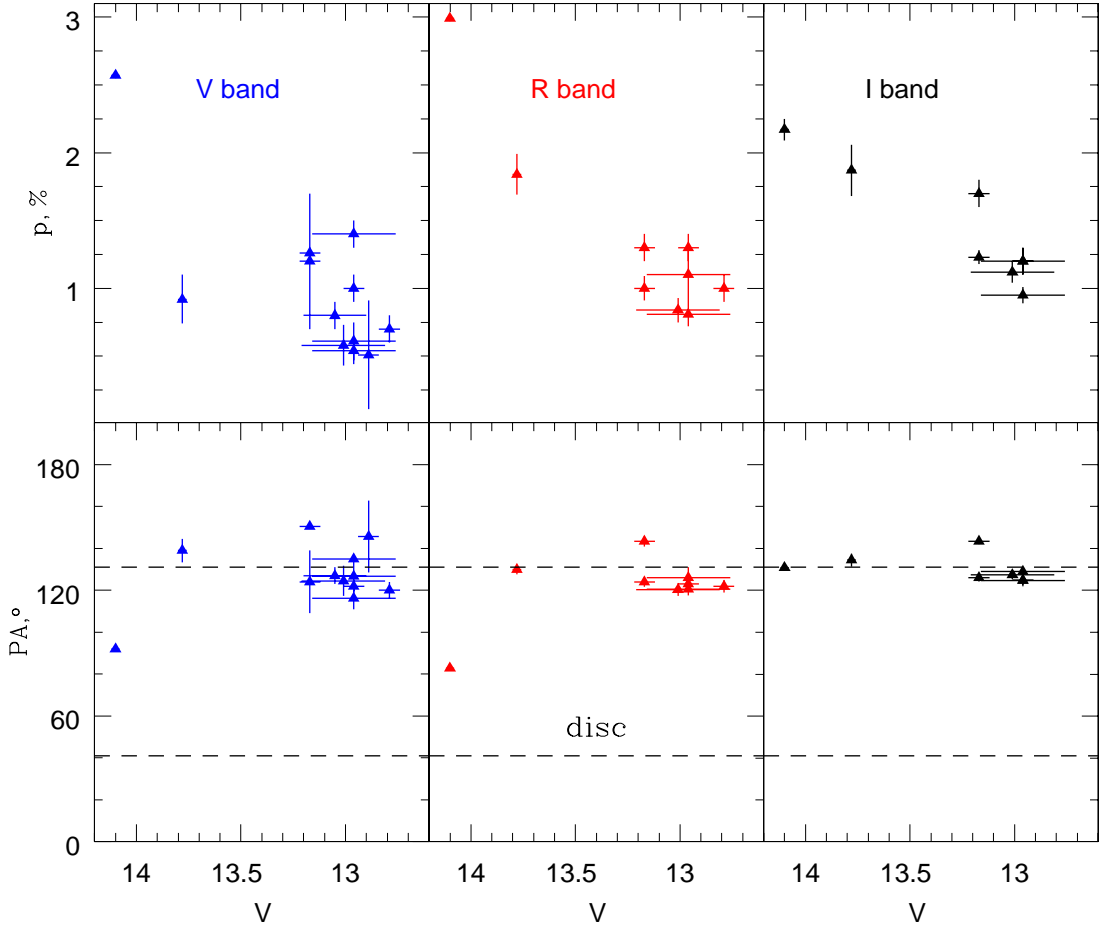}
 \end{center}
\caption{ The dependence of the degree (top panels) and the position angle PA
(bottom panels) of RW~Aur\,B's polarization on the $V$ magnitude.  The
left, middle and right columns  correspond to $V,$ $R$ and $I$ bands
respectively. The position angles, corresponding to PA of RW~Aur\,B's disc
and perpendicular to its direction are shown with the dashed lines. }
 \label{fig:VpPAs}
\end{figure*}

Our polarimetric observations indicate that when RW~Aur\,B fades, its optical flux becomes linearly polarized (up to 3 per cent in the $I$ band). The anti-correlation between the polarization and brightness was observed -- see the top panel of Fig.\,\ref{fig:VpPAs}. The position angle of polarization shows a significant erratic variability with a predominant direction across the major axis of the disc, as can be seen from the bottom panel of the figure.

			%%%%%%%%%%%%%  Fig: p,PA vs. V %%%%%%%%
%
%
			%%%%%%%%%%%%%%%%%%%%%
%
\subsection{Spectroscopy}
 \label{sect:spectroscopy}

\begin{figure}
 \begin{center}
\includegraphics[scale=0.45]{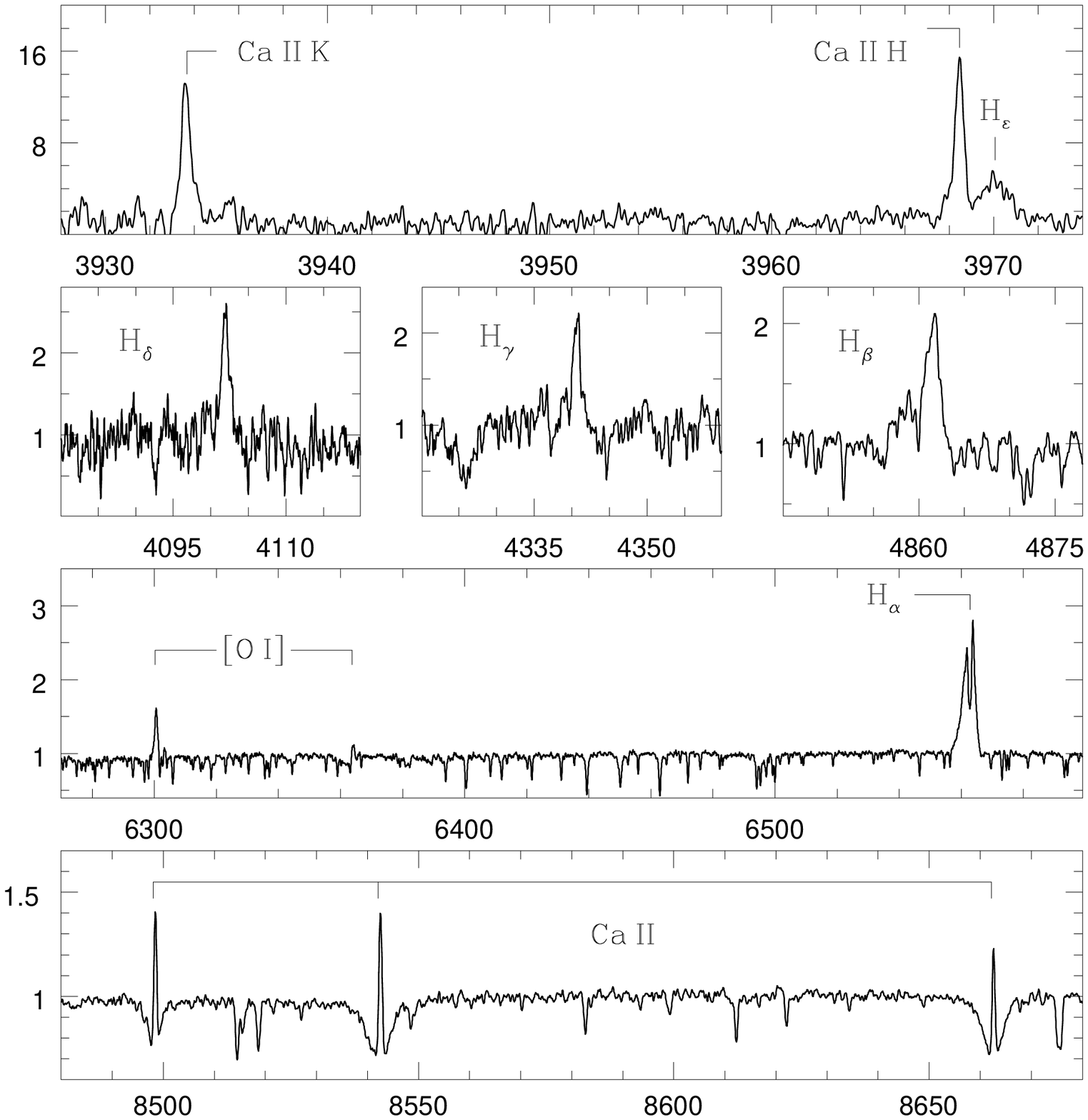}
 \end{center} 
\caption{Regions of RW~Aur\,B 2018 spectrum with emission lines. The monochromatic flux $F_\lambda$ along vertical axes is normalised to the average flux in 3940--3960~{\AA} band in the upper panel and to the continuum level in other panels.}
 \label{fig:emiss-low-res}
\end{figure}

  The three high resolution spectra of 1998, 2016 and 2018 (see Section~\ref{sect:observation}) show the stellar absorption lines and the emission lines of Hydrogen Balmer series (from H$\alpha$ to H$\varepsilon),$ \ion{Ca}{ii} (H, K, infrared triplet), \ion{He}{i}~$\lambda$5876\,{\AA} and [\ion{O}{i}] $\lambda$6300, 6363~{\AA} -- see Fig.\,\ref{fig:emiss-low-res} and Fig.\,\ref{fig:Na}.

			%%%%%%%%%%%%%  Fig: emiss-lines-low-res  %%%%%%%%
%
%
			%%%%%%%%%%%%%%%%%%%%%
 
Profiles and relative depths of absorption lines in all our spectra do not differ noticeably.
But the radial velocities found from the 2016 and 2018 spectra differ at $5\sigma$ level: $V_{\rm r} = 15.19 \pm 0.09$ and $14.53 \pm 0.08$\,km\,s${}^{-1},$ respectively. The accuracy of $V_{\rm r}$ in the 1998 spectrum is worse: $15.7 \pm 0.5$\,km\,s${}^{-1}.$ Note that \citet{White-Hillenbrand-2004} and \citet{Nguen-2012} found $V_{\rm r}=15.00 \pm 0.03$ and $15.9 \pm 0.5$\,km\,s${}^{-1},$ respectively. Thus, we believe that the measured radial velocity of RW~Aur\,B is variable with the amplitude $\sim0.5-1.0$\,km\,s${}^{-1}.$

Our measurements of the projected rotational velocity $v\sin\,i= 13.5 \pm 0.5$\,km\,s${}^{-1}$ are between values found by \citet{White-Hillenbrand-2004} and \citet{Nguen-2012} -- $12.2 \pm 1.6$ and $14.5 \pm 1.8$\,km\,s${}^{-1},$ respectively -- and are in agreement with both within the errors of measurements.

\begin{figure}
 \begin{center}
\includegraphics[scale=0.45]{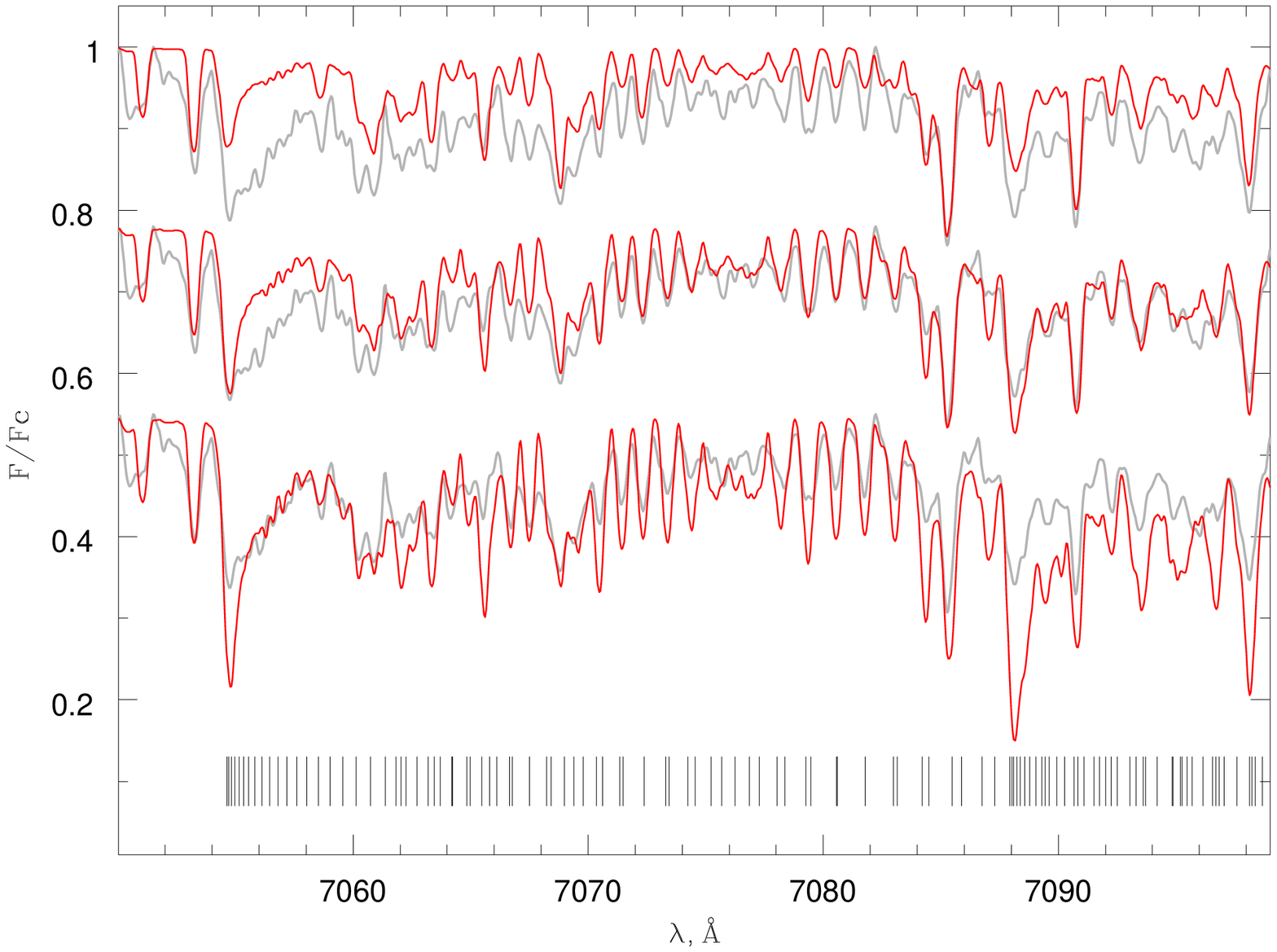}
 \end{center}
  \caption{Comparison of RW~Aur\,B 2016 spectrum (grey) in the vicinity of TiO $\gamma$-system (A${}^3{\rm \Phi}$ - X${}^3{\rm \Delta})$ molecular band with the {\sc sme} theoretical models (from top to bottom): $T_{\rm eff}=4200,$ 4100 and 4000 K (red lines). The solar composition \citep{Asplund-09} and $\log g=3.8$ were adopted. Spectra are shifted relative to each other for clarity. Position of TiO lines, shown at the bottom of the figure, were adopted from the {VALD} database \citep{VALD-15}.}
\label{fig:TiO}
\end{figure}

\begin{figure}
 \begin{center}
\includegraphics[scale=0.45]{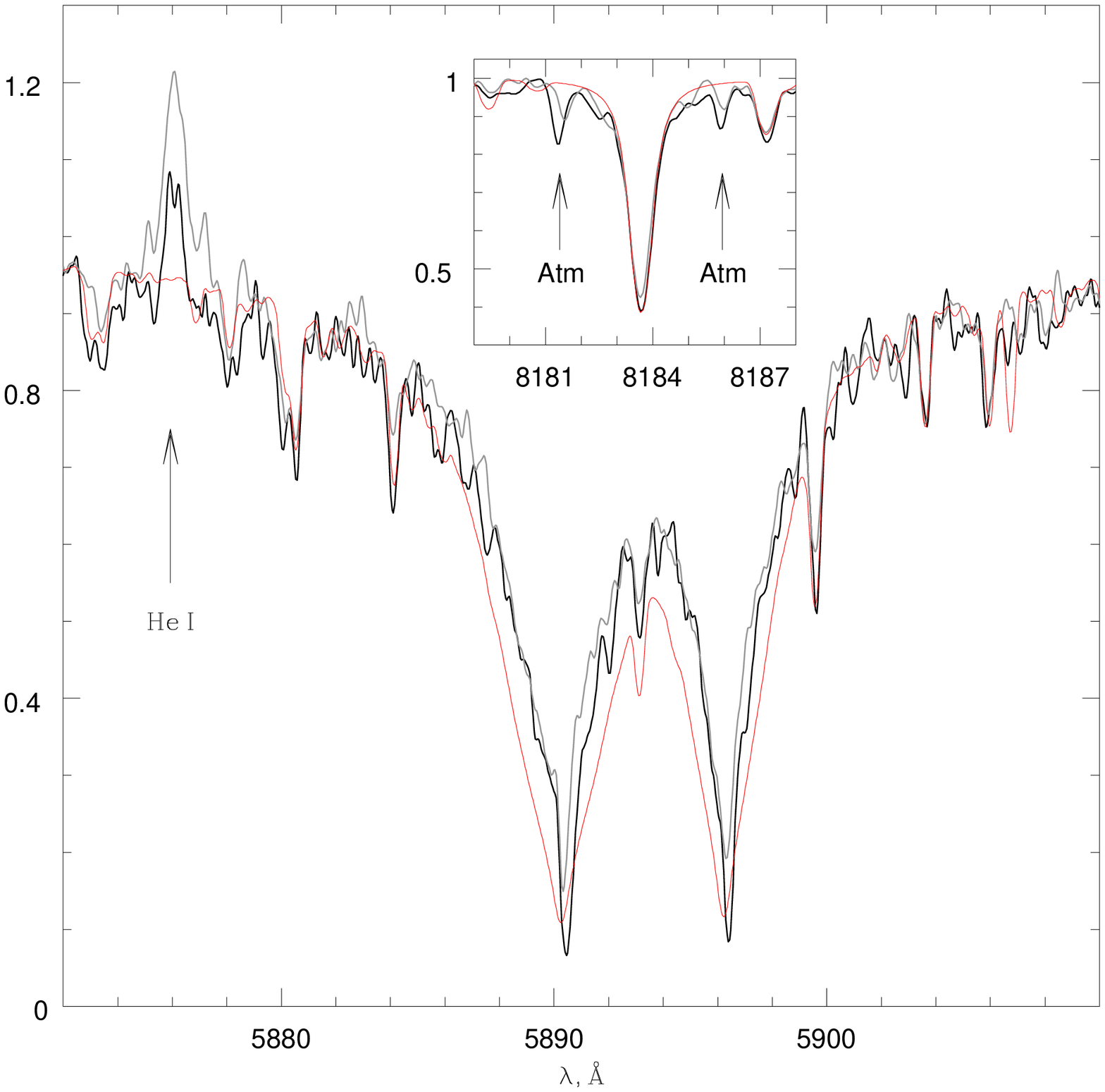}
 \end{center}
  \caption{Comparison of RW Aur B's 2016 (black line) and 2018 (grey line) spectra in the vicinity of the \ion{Na}{i} resonant doublet and subordinate 8183.26~{\AA} line (the insert) with $T_{\rm eff}=4200$~K, $\log \,g=3.8$ {\sc sme} theoretical model (red lines). Positions of \ion{He}{i} 5876 and some atmospheric H$_2$O lines are also shown.}
 \label{fig:Na}
\end{figure}

Comparison of the observed and {\sc sme}\footnote{\url{http://www.stsci.edu/~valenti/sme.html}} theoretical spectra  \citep{Piskunov16}, calculated for solar composition \citep{Asplund-09}, indicates that the effective temperature of the star $T_{\rm eff}$ is between 4100 and 4250~K -- see e.g. Fig.\,\ref{fig:TiO}.  Note that the depth of TiO lines shown in the figure depends only slightly on the surface gravity at $3.6 \leqslant \log g \leqslant 4.0.$ A possible presence of veiling ($r\lesssim0.1$ at $\lambda>7000$\,\AA) cannot affect significantly our estimate of $T_{\rm eff}.$

			%%%%%%%%% Fig.TiO %%%%%%%%%%%%
%
%
			%%%%%%%%%%%%%%%%%%%%%

Profiles of some lines of metals are much more sensitive to $\log g,$ and using them we found that $\log g=3.8 \pm 0.1.$ It can be seen from Fig.\,\ref{fig:Na} that theoretical profiles of the {\sc sme} model with $T_{\rm eff}=4200$\,K and $\log g=3.8$ describe well observed profile of \ion{Na}{i} $\lambda$8183.26\,{\AA} subordinate line as well as wings of the \ion{Na}{i} D resonant doublet lines. Some disagreement between calculated and observed profiles of the doublet lines in their central part we will discuss in the next Section.

				%%%%%%%%% Fig. NaD+subord %%%%%%%%%%%%
%
%
			%%%%%%%%%%%%%%%%%%%%%

				%%%%%%%%%%%%%%%%%%%%%%%%%%

\section{Interpretation}
\label{PandC-interp}
\subsection{RW Aur B as an UX Ori type star}\label{PandC-interp-1}

Photometric and polarimetric behaviour of RW~Aur\,B (e.g. the duration and amplitude of dimming events, the optical colour-magnitude diagrams, the anti-correlation between the brightness and polarization degree) resembles variability of UX~Ori type stars (UXORs). It is considered that these phenomena is connected with irregular eclipses of a young star by circumstellar dust, such as the star is surrounded by a disc, which scatters and polarizes a fraction of the stellar radiation toward the observer \citep{Grinin-88, Grinin-2001}.

As follows from Fig.\,\ref{fig:VpPAs}, PA of the polarization vector of RW~Aur\,B is oriented nearly perpendicular to the major axis of the on-sky projection of its disc that is expected for a disc scattering \citep{Whitney-93}. According to \citet{Oudmaijer-01} the variability of PA during the eclipse can be caused by chaotic changes in the illumination or/and visibility of a protoplanetary disc. In this regard, the star strongly differ from RW~Aur\,A, which has much larger amplitude of $p$ variability, and PA of its polarization is parallel to the major axis of its circumstellar disc that is best explained by scattering on dust particles of a dusty disc wind \citep{Dodin-18b}.

\subsection{SED, stellar and accretion parameters}\label{PandC-interp-2}
\begin{figure}
 \begin{center}
\includegraphics[scale=0.4]{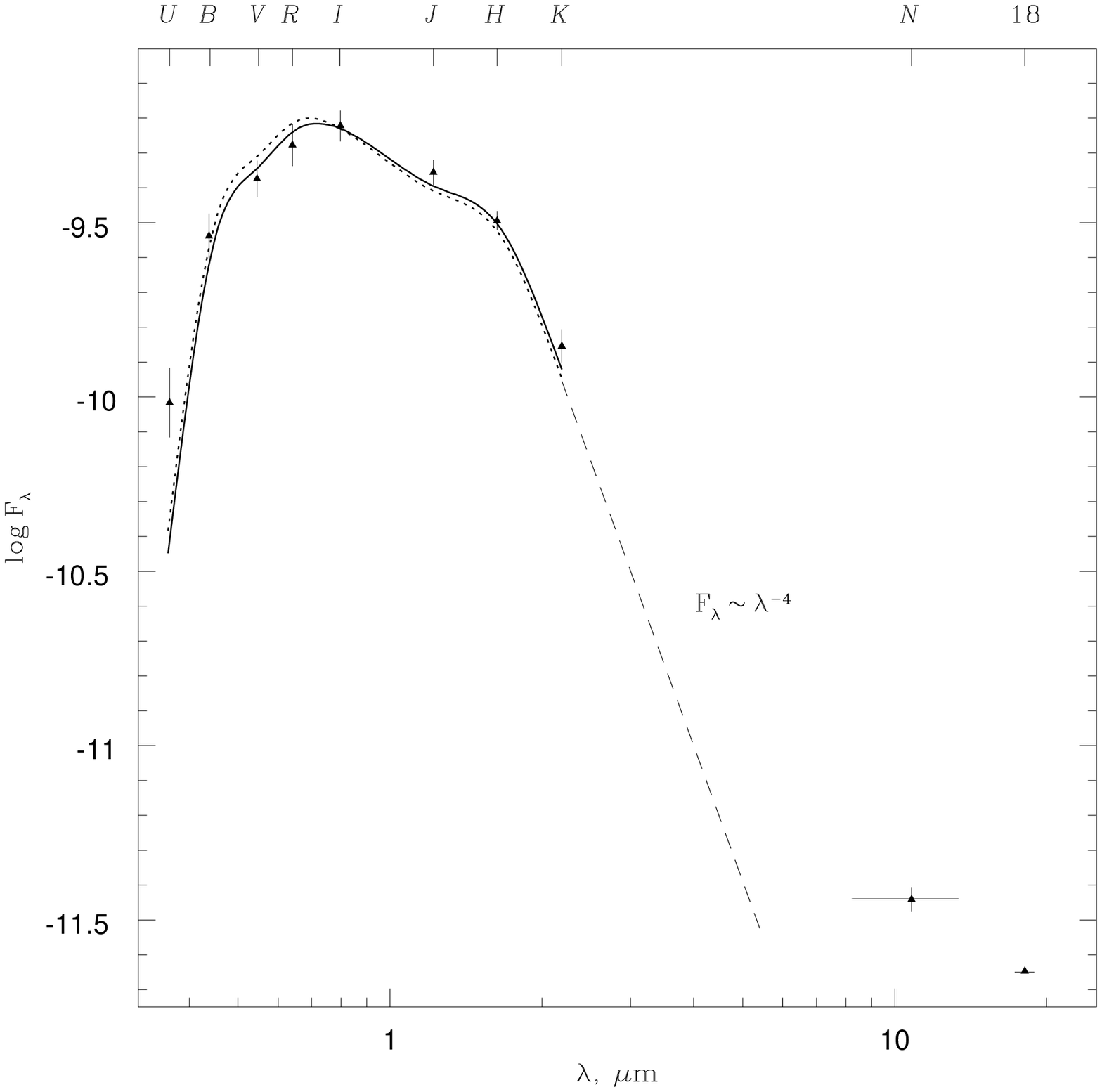}
 \end{center}
  \caption{Comparison of the dereddened SED of RW~Aur\,B (triangles) with the theoretical BT-settle models with $T_{\rm eff}=4100$~K and 4200~K (solid and dotted lines, respectively).  Dashed line represents the Rayleigh--Jeans approximation.  See text for details.  }
 \label{fig:sedB}
\end{figure}

The spectral energy distribution of RW~Aur\,B is shown in Fig.\,\ref{fig:sedB}. Black triangles in the figure with respective error bars correspond to the parameters found from the fit of our $UBVRIJHK$ data with Gaussian (Fig.\,\ref{fig:hist}),
and therefore they represent the bright state with its small-scale variability.  These data are supplemented by observations of \citet{McCabe-06} in the $N$ ($\lambda_{{\rm eff}}=10.8,$ $\Delta \lambda = 5.15$\,{\micron}) and $IHW18$ ($\lambda_{{\rm eff}}=18.1,$ $\Delta \lambda = 1.6$\,{\micron}) bands.  Solid and dotted curves represent a theoretical SED based on colour indexes in $0.35 < \lambda <2.2$~{\micron} spectral band calculated by \cite{Pecaut-Mamajek-2013} using solar abundance BT-settle models \citep{Allard-2012} with $T_{\rm eff}=4100$ K and 4200~K, respectively. Note that these theoretical curves are practically identical in the scale of the figure for $3.5 \leqslant \log g \leqslant 4.0.$ We found that just these curves reasonably well fit dereddened observations, if to use the interstellar reddening curve of \citet{Cardelli+1989} with $A_{\rm V}=0.65$ and $R_{\rm V}=3.1.$

			%%%%%%%%% Fig.SED_B %%%%%%%%%%%%
%
%
			%%%%%%%%%%%%%%%%%%%%%

Both models correspond to the same luminosity of the star $L_*\approx 0.59$~$L_\odot,$ which exactly coincides with the result of \citet{Herzeg-Hillebrandt-2014}, if to replace their distance to the star from 140 to 163 pc, in spite of significantly less $A_{\rm V}=0.1$ at the same $T_{\rm eff},$ which they found.

Our $A_{\rm V}=0.65,$ corresponding to the mean bright state of the star, is also larger than $A_{\rm V}=0.32 \pm 0.11$ found by \citet{Ghez+97}, as well as $A_{\rm V}$ for RW~Aur\,A before the 2010-2018 dimming events $(\approx 0.3$ according to \citealt{Petrov-01}). On the other hand, \citet{Gunther-18} concluded from 2013, 2015 and 2017 yr X-ray observations that the hydrogen column density in the direction of RW~Aur\,B is $N_{\rm H} = \left( 3 \pm 1 \right) \times 10^{21}$\,cm$^{-2}$ that corresponds to $0.9< A_{\rm V} <2.2,$ if to use the relation $N_{\rm H} = (2 \pm 0.2) \times 10^{21}\, A_{\rm V},$ appropriate for the IS medium \citep{Predehl-Schmitt-95}.  \citet{Andrews-2013} and \citet{Csepany-17} found even larger $A_{\rm V}$ values $(2.35 \pm 0.79$ and $3.3^{+0.3}_{-0.8},$ respectively) from SED of the star, but their results are based on non-simultaneous broad-band photometry.

Bearing in mind the uncertainty of RW~Aur\,B's average brightness (see Fig.\,\ref{fig:hist}) and possible presence of hot and cold spots on its surface, we will adopt the following parameters of the star: $L_*\approx 0.6 \pm 0.1$ $L_\odot$ and $T_{\rm eff}=4100-4200$~K.

\begin{figure}
 \begin{center}
\includegraphics[scale=0.65]{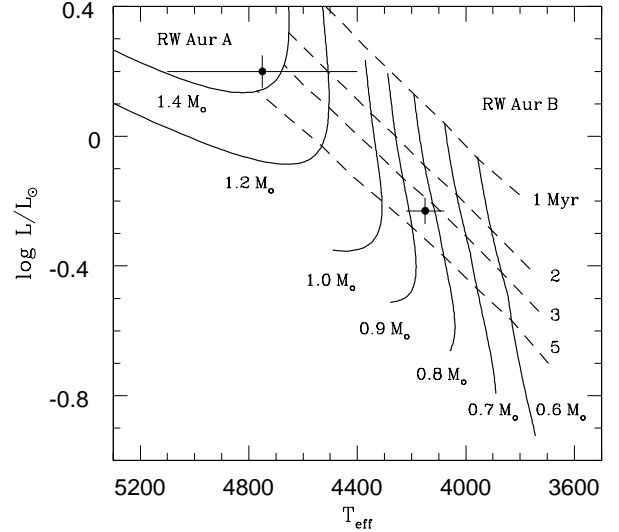}
 \end{center}
  \caption{Position of RW~Aur\,A and RW~Aur\,B in the HR diagram. The theoretical evolutionary tracks of pre-main sequence stars of different masses and the isochrones of different ages \citep{Baraffe+2015} are shown with the solid and dashed lines, respectively.}
 \label{fig:age}
\end{figure}

The position of the star in the Hertzsprung--Russell (HR) diagram, corresponding to these parameters, is shown in Fig.\,\ref{fig:age} along with the theoretical evolutionary tracks and isochrones of \citet{Baraffe+2015}.  It follows from the figure that the mass of RW~Aur\,B is $M_*=0.85 \pm 0.05$ M$_\odot.$ This value can be compared with estimations found with a similar approach by \citet{White-Ghez-01} -- $0.93 \pm 0.09$ M$_\odot$ and \citet{Kraus-2009} -- $0.86\pm 0.10$ M$\sun.$ Note also that \citet{Rodriguez-18} found 0.5 M$_\odot < M_* < 1 M_\odot$ from an analysis of the position-velocity diagram of RW~Aur\,B's circumstellar disc.

			%%%%%%%%% Fig.HR-diagram %%%%%%%%%%%%
%
%
			%%%%%%%%%%%%%%%%%%%%%

With adopted values of $L_*$ and $T_{\rm eff}$ the radius of the star is $R_* = 1.5 \pm 0.1$ $R_\odot,$ the free-fall velocity is $V_\infty = 470$\,km\,s$^{-1}$ and the surface gravity $\log g\approx 4.0,$ which is in a reasonable agreement with our estimate from the spectral data $\log g=3.8 \pm 0.1$ (Section~\ref{sect:spectroscopy}).

A number of authors noted that the stellar spectrum is somewhat veiled: \citet{White-Hillenbrand-2004} found $r=0.26 \pm 0.12$ and $0.17 \pm 0.13$ near $\lambda=0.65$ and 0.84~{\micron}, respectively, \cite{Herzeg-Hillebrandt-2014} found $r=0.18$ at $\lambda=0.751$~{\micron} and \citet{Edwards-2006} found $r=0.1$ in the $Y$ band.  As can be seen from Fig.\,\ref{fig:sedB}, an excess emission is definitely present in the blue part of RW~Aur\,B spectrum, where the S/N ratio in our spectra is poor, but one can estimate the contribution of the excess emission in this region from the broad band photometry. Using the theoretical {\sc sme} model as a template, we found from our RW~Aur\,B's optical spectra that $r$ does not exceed 0.2 in the $V$ band, $r<0.15$ in the $R$ band and $r<0.1$ longward $\lambda \approx 0.7$~{\micron} -- see the insert of Fig.\,\ref{fig:Na}.  We thus conclude that the relative contribution of the excess emission in the optical band, which presumably originate in the hot (accretion) spot, does not exceed 20 per cent of $L_{*},$ i.e. the accretion luminosity of the star is $L_{\rm acc}< 0.1$\,$L_\odot,$ in agreement with the value $0.05$~$L_\odot$ found by \citet{Eisner-07}.  Then from the relation $L_{\rm acc} \approx GM_*\dot M_{\rm acc}/R_*$ we obtain that the accretion rate $\dot M_{\rm acc}$ is $\lesssim 5\times 10^{-9}$ M$_\odot$ yr$^{-1}$ during our observations that is an order of magnitude less than the value found by \citet{White-Hillenbrand-2004}.

It is reasonable to explain the small-amplitude photometric
variability of the star in the optical band, which is described by the
respective Gaussians in Fig.\,\ref{fig:hist}, by small-scale dimmings,
cold spots,  and a rotational modulation related to the presence of a hot spot as well as by a non-stationary accretion process.  It could explain why our spectra are significantly less veiled than the spectrum observed by \citet{White-Hillenbrand-2004}: in particular, the EW of \ion{Li}{i} $\lambda$6707.9\,{\AA} line in our 2016 and 2018 spectra is $0.58 \pm 0.01$~\AA{}, it is larger than the value $0.50 \pm 0.02$~\AA{} found by these authors from the spectrum, in which the EW of H$\alpha$ emission line was 4-5 times larger than in our spectra.

We did not find the rotational period of RW~Aur\,B from the photometric observations, possibly due to a superposition of non-periodic dimming events. But one can estimate the period from our data using the following relation:
\begin{equation} 
P_{\rm rot} = \frac{2\pi R_*}{v\sin i}\times \sin i \approx
5.5 \times \sin i, \,\,\, \mbox{ days}.  
\end{equation} 
It is not obvious that an inclination angle $i$ of the rotation axis of the star to the line of sight is the same as that of the disc, i.e. $i\approx 70\degr$ \citep{Rodriguez-18, Manara-2019}.

It follows from Fig.\,\ref{fig:sedB} that theoretical models predict a significantly lower flux than observed not only in the blue part of the spectrum, but also at $\lambda>2$~{\micron}. We suppose that the NIR excess is due to the contribution of RW~Aur\,B's circumstellar disc discovered from Submillimeter Array \citep{Andrews-2013} and ALMA \citep{Rodriguez-18} interferometric observations.  We suppose that the significant and variable contribution of the disc emission in the $K$ band explains why the photometric behaviour of the star in this and the $UBVRIJH$ bands is different as follows from Fig.\,\ref{fig:hist}.  Indeed, as follows from Fig.\,\ref{fig:colors} the less brightness of the star in the $K$ band the closer colour indexes $J-K$ and $H-K$ to the respective colours of theoretical BT-Settle models with $T_{eff}=4100-4200$\,K \citep{Pecaut-Mamajek-2013}. Probably the variability of disc emission in the $K$ band is due to a non-stationary character of the accretion.

\subsection{Spectral manifestations of accretion}\label{PandC-interp-3}
Consider now what kind of information one can extract from analysis of the emission lines in RW~Aur\,B spectra. We will use below the term {\lq}residual profiles{\rq}, meaning the difference between observed and theoretical {\sc sme} ($T_{\rm eff}=4200$~K, $\log g =3.8,$ solar abundances) profiles $F_\lambda (\lambda)$ normalised to the continuum level.

\begin{figure}
 \begin{center}
\includegraphics[scale=0.6]{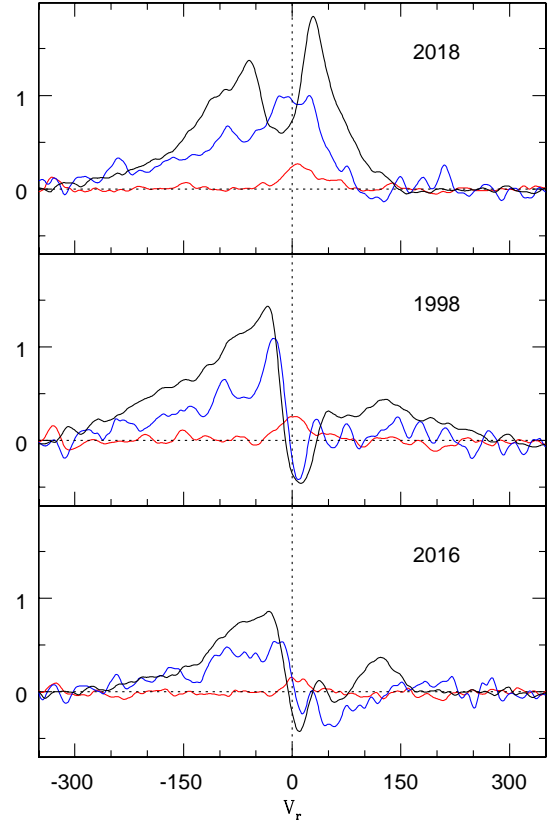}
 \end{center}
  \caption{Residual profiles of H$\alpha,$ H$\beta$ and \ion{He}{i} $\lambda$5876\,{\AA} lines ( black, blue and red curves, respectively) in 1998, 2016 and 2018 spectra of RW~Aur\,B.
  }
 \label{fig:HHe-profs}
\end{figure}

Let us discuss initially the H$\alpha,$ H$\beta$ and \ion{He}{i}~$\lambda$5876\,{\AA} lines, which have asymmetric and strongly variable profiles.  We arranged the residual profiles of these lines in Fig.\,\ref{fig:HHe-profs} in descending order of the intensity of the H$\alpha$ line from the top to bottom panels, such as EW of the most strong H$\alpha$ line (2016 spectrum) is 4.0\,{\AA}.  The following features of the lines can be noted.
\begin{enumerate}
\item  The red wing of the \ion{He}{i} line observed in 2018 and 1998 extends to $\approx +80$\,km\,s$^{-1},$ while the blue one extends to $\approx -50$\,km\,s$^{-1}$ only. In the 2016 spectrum the line is more narrow and less intensive. In all cases the peak of the line is redshifted by a few km\,s$^{-1}.$ These types of profiles match the theoretical expectations in the model of magnetospheric accretion \citep{Lamzin-1998, Dodin-18}.
\item  The profiles of the H$\alpha$ and H$\beta$ lines are similar to each other. These profiles are much wider and have a more complicated structure than those of the \ion{He}{i}~$\lambda$5876\,{\AA} line.  The blue wings of the hydrogen lines extend up to $\approx -300$\,km\,s$^{-1}$ in all our spectra as well as in the case of the H$\alpha$ line observed by \citet{White-Hillenbrand-2004} and the \ion{He}{i} $\lambda$1.083~\,{\micron} line observed by \citet{Edwards-2006}.  It well can be that this part of the hydrogen lines is originated in the magnetospheric wind.
\item  Extension of the red wing of the hydrogen lines varies from $\approx +300$\,km\,s$^{-1}$ (in our 1998 spectrum as well as in the case of H$\alpha$ line in the spectrum of \citealt{White-Hillenbrand-2004}) to $\approx +150$\,km\,s$^{-1}.$ Apparently, there are two distinct absorption features in the profiles of hydrogen lines. 

The centre of the first feature changes its position $V_1$ from $\approx +15$\,km\,s$^{-1}$ in the 1998 and 2016 spectra to $\approx -20$\,km\,s$^{-1}$ in the 2018 spectrum.  A similar absorption feature was also observed in the H$\alpha$ line at $V_1 \approx -45$\,km\,s$^{-1}$ by \citet{White-Hillenbrand-2004} as well as in the \ion{He}{i} $\lambda$1.083~\,{\micron} line at $V_1 \approx -30$\,km\,s$^{-1}$ by \citet{Edwards-2006}.  The absorption falls below the continuum level at $V_1>0$ that means that at these moments the formation region of the feature is projected to the star. Hence, we conclude that the feature is associated with a magnetospheric accretion flow, especially since negative values of $V_1$ are also possible in this case -- see Figure\,5 in \citet{Errico-2001} and Figure\,17 in \citet{Romanova-2003}.

The second absorption feature centred at $V_2 \approx +50$\,km\,s$^{-1}$ can be seen in our 1998 and 2016 spectra.  In the last case, it apparently falls below the continuum level, so we also conclude that it is also associated with an accretion flow, but closer to the star as far as $V_2 > V_1.$
\end{enumerate}

It is not possible to present a more detailed model of the gas flow in the vicinity of RW~Aur\,B basing on the available spectra. We can only assume that the magnetic axis of the star is inclined to its rotation axis and the observed variations of hydrogen and helium lines profiles are a result of rotational modulation. Probably, a non-stationary character of the accretion or/and the geometry/strength of the stellar magnetic field are also important.

			%%%%%%%%% Fig. Comparison of H_a, H_b & He profiles %%%%%%%%%%%%
%
%
			%%%%%%%%%%%%%%%%%%%%%

\begin{figure}
 \begin{center}
\includegraphics[scale=0.45]{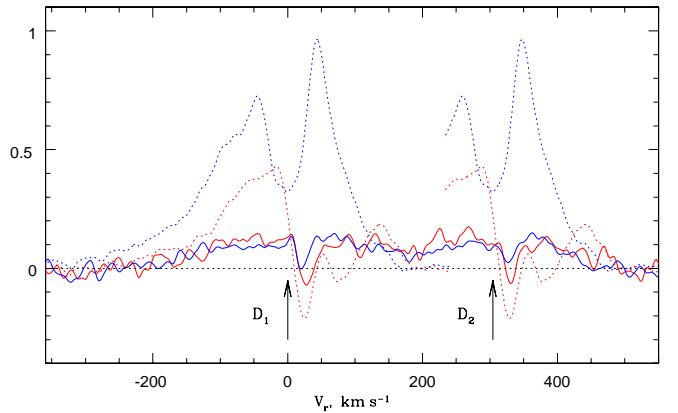}
 \end{center}
  \caption{Comparison of the residual profiles of H$\alpha$ (dotted curves) and \ion{Na}{i} D (solid curves) lines in 2016 (red) and 2018 (blue) spectra of RW~Aur\,B. For ease of comparison the profiles of H$_\alpha$ line are compressed two times. Vertical arrows mark stellar rest velocity of the sodium doublet lines.  See text for details. }
 \label{fig:cmpNaHa}
\end{figure}

We found that the residual profiles of \ion{Na}{i}~D lines are broad emission peaks in the central part of the respective absorption lines with superimposed sodium interstellar absorption features.  It appears that profiles of these emission features vary in the same way as the profiles of H$\alpha$ line, as can be seen in Fig.\,\ref{fig:cmpNaHa}.  As far as the sodium D$_1$ and D$_2$ lines overlap each other, we plot in the figure two profiles of H$\alpha$ line shifted by $\Delta \lambda=\lambda_{\rm D2}-\lambda_{\rm D1}\approx5.97$\,{\AA} (or 304\,km\,s$^{-1}$) relative to each other.  It gives possibility to see that both the blue wing of the D$_2$ line and the red wing of the D$_1$ line correspond to the wings of H$\alpha$ line.  We conclude therefore that the emission components of the \ion{Na}{i}~D lines as well as hydrogen and helium lines are connected with the accretion process. Note that the \ion{Na}{i}~D lines are routinely have broad emission components in CTTSs including RW~Aur\,A -- see e.g.  \citet{Alencar-05,Takami-16}.

			%%%%%%%%% Fig. Compar Ha & Na D profiles %%%%%%%%%%%%
%
%
			%%%%%%%%%%%%%%%%%%%%%
\subsection{Origin of the forbidden lines}\label{PandC-interp-4}
\citet{White-Hillenbrand-2004} observed emission forbidden lines of \ion{O}{i}, \ion{N}{ii} and \ion{S}{ii} in RW~Aur\,B spectrum. The [\ion{O}{i}] and [\ion{N}{ii}] lines have narrow profiles centred at the stellar rest velocity, whereas profiles of the [\ion{S}{ii}] $\lambda6716,$ 6731\,{\AA} lines are redshifted by $\approx 100$\,km\,s$^{-1}$ and significantly more broad. We, as well as \citet{Duchene-99} and \citet{Petrov-15}, found only [\ion{O}{i}] $\lambda$6300 and 6363\,{\AA} lines (see the bottom panel of Fig.\,\ref{fig:emiss-low-res}), however the first of these lines is out of spectral orders in the 1998 spectrum.

Weak emission features in the vicinity of the [\ion{S}{ii}] lines are also present in our spectra, but their profiles are exactly the same as profiles of the respective strong lines of RW~Aur\,A counter jet observed at the same dates \citep{Takami-16}. RW~Aur\,B is significantly closer to the red lobe of RW~Aur\,A's jet than to its blue one \citep[see e.g. Fig.1 of][]{Dougados-00}, so we conclude that the redshifted [\ion{S}{ii}] lines in our RW~Aur\,B spectra appear due to insufficient subtraction of the counter jet contribution.  As far as high velocity components of forbidden lines originate in jets \citep{Nisini-2018} and RW~Aur\,B apparently has no such jet \citep{Dougados-00}, we suppose that the [\ion{S}{ii}] lines in \citet{White-Hillenbrand-2004} spectra have the same origin.

The [\ion{O}{i}] $\lambda$6300\,{\AA} line in all our spectra has a practically constant Gaussian-like profile centred at the stellar rest velocity as well as in Figure 4 of \citet{Petrov-15}.  We found that EW of the line is $\approx 0.8$\,{\AA} with an accuracy about 10 per cent.  As follows from Fig.\,\ref{fig:sedB} the dereddened continuum flux at 0.63\,{\micron} is $\approx 5.7\times 10^{-14}$\,erg\,s$^{-1}$\,cm$^{-2}$\,\AA$^{-1}$ therefore the luminosity in the line is $\approx 1.4\times 10^{29}$\,erg\,s$^{-1}.$ 

The FWHM of the [\ion{O}{i}] $\lambda$6300\,{\AA} line is $\approx 45$\,km\,s$^{-1},$ so, according to the classification by \citet{Nisini-2018}, the line has the so-called low-velocity component (LVC) only, which presumably originates in a disc wind.  These authors found that the luminosity in this line $L_{\rm LVC}$ is well correlated with the accretion luminosity $L_{\rm acc},$ 
and it turned out that our respective values satisfy this relation within its accuracy, additionally confirming self-consistency of our parameters of RW~Aur\,B. The absence of the [\ion{O}{i}] $\lambda=5577$\,{\AA} line $F_{\rm 5577}/ F_{\rm 6300} < 0.05$ indicates that the gas temperature in the disc wind is $<1.2\times 10^4$\,K \citep{Nebulio-2015}.

\subsection{Chromospheric activity}\label{PandC-interp-5}
\begin{figure}
 \begin{center}
\includegraphics[scale=0.45]{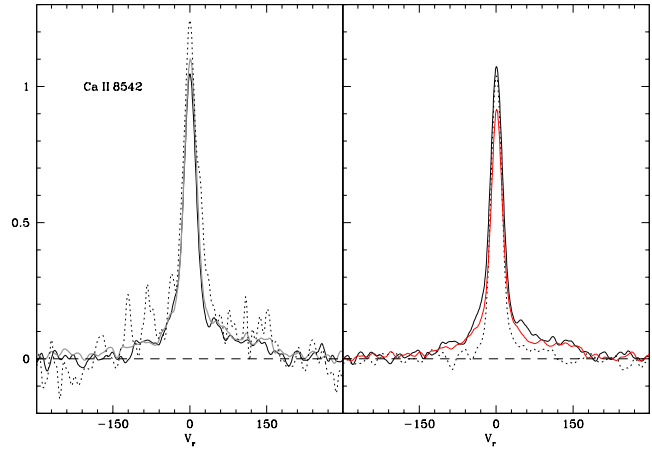}
 \end{center}
  \caption{Residual profiles of \ion{Ca}{ii} $\lambda = 8542$~{\AA} line in 1998, 2016 and 2018 spectra of RW~Aur\,B are shown in the left panel with dotted, solid and grey curves, respectively.  Comparison of the line profiles at $\lambda = 8498,$ 8542 and 8662\,{\AA} averaged over 2016 and 2018 spectra are shown in the right panel with dotted, solid and red curves, respectively.  }
 \label{fig:CaIRs}
\end{figure}

The emission components of the \ion{Ca}{ii} infrared triplet (IRT) lines $\lambda=8498,$ 8542 and 8662\,{\AA} in RW~Aur\,B spectrum differ significantly from the Balmer and \ion{He}{i}~$\lambda$5876\,{\AA} emission lines. To demonstrate this, we plot the residual profile of the $\lambda=8542$\,{\AA} line (the strongest line of IRT) in the left panel of Fig.\,\ref{fig:CaIRs}. One can see that the line profile is symmetric relative to the stellar rest frame, and the line flux does not vary significantly at least between the 2016 and 2018 spectra, in spite of the large variations in the hydrogen and helium lines (see Fig.\,\ref{fig:HHe-profs}). It appears that fluxes and profiles of the IRT emission components at $\lambda=8498$ and 8662\,{\AA} in the 2016 and 2018 spectra do not differ within the error of measurements, so we averaged the profiles of all IRT lines in these spectra and plotted the resulting profiles in the right panel of Fig.\,\ref{fig:CaIRs} (Our 1998 spectrum is more noisy and $\lambda=8498$ and 8584\,{\AA} lines are close to the end of the spectral order, so we exclude it from the analysis.) Now one can definitely state that all of the IRT lines have symmetric and much narrower profiles than the hydrogen and helium lines. The signal-to-noise ratio of our spectra in the vicinity of the \ion{Ca}{ii} H, K lines is very low, nevertheless we can state that these lines also look narrow and symmetrical relative to the stellar rest frame.

			%%%%%%%%% Fig. Residual Ca II IR lines profiles %%%%%%%%%%%%
%
%
			%%%%%%%%%%%%%%%%%%%%%

We suppose that the IRT lines as well as the H and K lines of \ion{Ca}{ii} originates from the chromosphere of RW~Aur\,B. We used the equation (12) of \citet{Busa-2007} to estimate the commonly used chromospheric activity index $R^{\prime}_{\rm HK}$ (the chromospheric or residual emission flux in the \ion{Ca}{ii} H+K lines, normalized to the bolometric flux) from the summed EW of all three IRT lines $\Delta W_{\rm IRT}.$ We found from the right panel of Fig.\,\ref{fig:CaIRs} that EWs of $\lambda=8498,$ 8584 and 8662\,{\AA} lines are 0.89, 1.62 and 1.31\,{\AA}, respectively, so $\Delta W_{\rm IRT} \approx 3.8$\,{\AA} and $\log R^{\prime}_{\rm HK} \approx -3.5.$

This $R^{\prime}_{\rm HK}$ value corresponds to stars with the most active chromospheres -- see e.g. Figure 5 in \citet{Busa-2007}\footnote{Note that \citet{White-Hillenbrand-2004} found even larger EWs of \ion{Ca}{ii} IRT lines: $EW_{8498}=5.9$ and $EW_{8662}=3.5$\,{\AA}.}.
It looks quite reasonable because the coronal X-ray luminosity of RW~Aur\,B also corresponds to a high level of activity: $\log L_{\rm X} \approx 30.2 \pm 0.1$ \citep{Skinner-2014,Schneider-15,Gunther-18}, i.e.  $\log L_{\rm X}/L_{\rm bol} \approx -3.2$ -- see e.g. Figure 5 in \citet{Preibisch-2005} for comparison.

The high level of chromospheric-coronal activity means that the star has a strong enough magnetic field -- see e.g.  \citet{Linsky-2017} and references therein. In turn, the magnetic field and hot or/and cold spots can significantly alter the EWs of absorption lines in RW~Aur\,B spectrum at $\log R^{\prime}_{\rm HK}> -4$ \citep{Spina-2020}, so we suppose that the accuracy of our estimate of the effective temperature of the star $T_{\rm eff}=4100-4200$\,K is quite reasonable.

According to \cite{Gunther-18} the temperature $T_2$ and the emission measure EM$_2$ of the coronal region, where the hard X-rays of RW~Aur\,B originate, are near $2\times 10^7$\,K and $10^{53}$\,cm$^{-3},$ respectively. Then one can find the scale height in this region $H_{\rm p} = \Re T_2/\mu g \approx 3\times 10^{11}$\,cm $\approx 2.5$\,R$_*.$ Here $\Re\approx 8.3\times 10^7$\,erg K$^{-1}$\,g$^{-1}$ and $\mu \approx 0.6$ are the universal gas constant and the mean molecular weight of the coronal gas, respectively.   According to \citet{Edwards-2006} the extension of the redshifted absorption feature of the \ion{He}{i} $\lambda$1.083\,{\micron} line is 400\,km\,s$^{-1}.$  At the stellar parameters of RW Aur B it corresponds to the free fall velocity from the height of $3R_*$ (or disc truncation radius 4$R_*),$ which is $\approx H_p$. It supports the viewpoint that the hot coronal gas $(T=T_2)$ fills all magnetosphere bounded by the accretion flow \citep{Lamzin-1999}.

RW~Aur\,B is a variable X-ray source, most likely due to coronal flares \citep{Skinner-2014, Gunther-18}. \citet{Furtig64} observed the flare in the optical band in one night of the unresolved monitoring of RW~Aur\,AB -- see the respective footnote in \citet{Herbig-Rao-1972}.  We did not observe such flares in RW~Aur\,B, and data of \citet{Kiliachkov-Shevchenko-1976, Worden-1981, Gahm-1990, Costigan-2014} demonstrate that it well could be a flare in RW~Aur\,A.

As was shown above, the rotational period of the star $P_{\rm rot}$ is significantly less than 10 days. This indicates that the chromospheric-coronal activity of the star is in the so-called saturated regime, when all the surface of the star is covered with active regions -- see e.g. \citet{Gudel-2004} and reference therein.  Probably, it explains why \ion{Ca}{ii} IRT lines have symmetric profiles in all our spectra.

\subsection{The binary system}\label{PandC-interp-6}
  After the close approach of two stars, which occurred about 400 years ago, the accretion rate of RW~Aur\, A becomes $\sim10^{-7}$ M$_\odot$~yr$^{-1}$   
\citep{Dodin-18b}, which is more than an order of magnitude larger than   
$\dot M_{\rm acc}$ of RW~Aur\, B (see above). It is not clear whether such different levels of activity of the binary components can be explained only by the fact that the mass of the primary and its disc are about 1.5 and $4-5$ times, respectively, larger than those of the secondary --
see Fig.\,\ref{fig:sedB} and \citep{Akeson-2019}.  But in any case, it is necessary to explain why RW~Aur\,A possess a powerful jet, while RW~Aur\,B shows no evidence for a jet. Although a discussion these issues is beyond the scope of our paper, we would like to pay attention to the following.

  It follows from 3D MHD simulations of \citet{Sheikhnezami-Fendt-2018} that jet launching from discs in binary systems
occurs only if an angle between planes of the disc and the orbit does not exceed  some critical value $\approx 10\degr,$ {\lq}beyond which a jet cannot persistently be formed out of a disc wind{\rq}.  \cite{Wurster-Bate-2019} also  
found that bipolar outflows are launched only in those models that have strong magnetic fields, which are initially parallel to the
rotation axis.  It supports the idea of \cite{Fendt-Zinnecker-98} that {\lq}a certain degree of axisymmetry as an essential ingredient for the jet
launching{\rq}.  Remind in this regard that the discs of RW~Aur\,A and B are 
misaligned \citep{Rodriguez-18,Manara-2019}, and perhaps this explains why the the primary has the jet, while the secondary does not. 

We plot in the HR diagram (Fig.\,\ref{fig:age}) the position of RW~Aur\,A, taking its parameters from the paper of \citet{Dodin-18b}.  It can be seen that our data at least do not contradict to the hypothesis that both stars have the same age of $3 \pm 1$ Myr\footnote{ \citet{Ghez+97} concluded that RW~Aur\,A is significantly younger than RW~Aur\,B, but \citet{White-Ghez-01} and \citet{Kraus-2009} found that the stars have the same age within the errors of measurements.}
, and thus RW~Aur\,AB {\lq}is a bona fide binary, rather than a one off encounter{\rq} \citep{Rodriguez-18}.

There is one more argument in favour of this conclusion. According to \citet{Berdnikov-17} the age of HH\,229 jet of RW~Aur\,A is $\sim 400$\,yr. But \citet{McGroarty-Ray-04} found one more HH object (HH\,835) $5.37\arcmin$ to the northwest of RW~Aur\,A on the axis of the HH\,229 outflow.  There is a  region of $4.6\arcmin$ between the end of the redshifted lobe of HH\,229 jet and HH\,835 without optical evidence for outflow activity. If HH\,835 is the result of outflow activity occurred during the previous fly-by of RW~Aur\,B, then the orbital period is the difference of the dynamical times of HH\,835 and the leading knot of HH\,229 flow.  According to \citet*{McGroarty+07} it is about $1000-1500$ yr, in agreement with the estimate of \citet{Bisikalo-12}.

			%%%%%%%%%%%%%%%%%%%%%
%
\section{Summary}
\label{summary-sect}

  Based on the results of our photometric, polarimetric and spectral observations, we concluded that RW~Aur\,B is a young star (the age is $3\pm 1$ Myr) with the effective temperature $T_{\rm eff} = 4100-4200$\,K, luminosity $L_*\approx 0.60$\,$L_\odot,$ radius $R_* \approx 1.5$\,$R_\odot$ and mass $M\approx0.85$\,M$_\odot.$ We also found that the extinction in the direction of the star is variable, with the median value of $A_{\rm V}=0.65.$ The star accrets matter from its circumstellar disc with an accretion rate $\dot M \lesssim 5\times 10^{-9}$\,M$_\odot$\,yr$^{-1}$ and an accretion luminosity $L_{\rm acc}\lesssim 0.1$~L$_\odot.$ A hot accretion spot on the stellar surface is responsible for the veiling of the spectra of the star at $\lambda \lesssim 2$\,{\micron}. At larger wavelengths the radiation from the accretion disc dominates in the SED of RW~Aur\,B.

  The radial velocity of the star is near 15\,km\,s$^{-1}$ and varies with an amplitude of $\sim 0.5 - 1.0$\,km\,s$^{-1},$ possibly due to a change of the hot spot position relative to the Earth when the star rotates. The projected rotational velocity of the star is $v\sin\,i \approx 13.5$\,km\,s${}^{-1},$ and its rotational period is $<5.5^{\rm d}.$

There is no apparent regularity in the photometric variability of the star, and
the brightness variations are well correlated in all bands except $K.$ 
The magnitude distributions in the visual and $J,$ $H$ bands can mostly be described by Gaussian curves, $\sigma$ of which has a maximum value of $0.25^{\rm m}$ in the $U$ band and a minimum value of $0.07^{\rm m}$ in the $H$ band. Probably there are a number of reasons for this phenomenon including rotational modulation of the hot spot emission and its variability due to fluctuations of the accretion rate as well as small scale obscurations by the circumstellar dust.
Presumably the photometric variability in the $K$ band is connected with a variable contribution of the accretion disc.

We present arguments in favour of the accretion origin for the emission in \ion{He}{i}~$\lambda$5876\,{\AA} and \ion{Na}{i}~D lines as well as in the red part of profiles of hydrogen Balmer lines. It well can be that the emission in the blue part of the Balmer lines profiles originates in a magnetospheric wind, but we can not state this for
sure.  We limit ourselves to these qualitative statements, because more
spectra observed at different rotational phases are required to make quantitative conclusions about the geometry and physical parameters of the gas flow in the vicinity of the star.

We concluded that emission components of \ion{Ca}{ii} H, K and IRT lines in RW~Aur\,B spectrum originate in the chromosphere of the star.  The index of its chromospheric activity is high $\log R^{\prime}_{\rm HK} \approx -3.5$  as well as the ratio $\log L_{\rm X}/L_{\rm bol} \approx -3.2.$ 

We also observed irregular short-term $(\sim$ days) dimmings of the star in the visual band with an amplitude $\Delta V$ up to $1.3^{\rm m}.$ The fainter the star during these events the redder it is, so we believe that the dimmings occur due to eclipses of the star by circumstellar dust clouds of unknown origin.  The dimmings are accompanied with an increase in the linear polarization up to 3 per cent in the $I$ band. Apparently RW~Aur\,B is a UX~Ori type star and the observed polarization is the result of scattering of the stellar radiation by dust in the circumstellar disc.

In this respect the star differs from RW~Aur\,A in its dim state 2010-2019, the polarization of which is due to the scattering of stellar light in the dusty disc wind.  RW~Aur\,B also has a disc wind, judging by the presence of the forbidden \ion{O}{i} lines in its spectra, but we have no reason to believe that dust clouds in the disc wind are responsible for the observed dimmings of the star.

The luminosity $L_{\rm OI}$ of the [\ion{O}{i}] $\lambda$6300 line in RW~Aur\,B spectra is quite consistent with the statistical dependency $L_{\rm OI}$ vs.  $L_{\rm acc}$ for CTTSs, indicating that the disc wind of RW~Aur\,B is not unusually strong.

We note that the close approach of A and B components of RW~Aur binary much stronger affected the disc structure of the A than B component.  We suggest that this may be in large part due to differences in the  inclinations of the circumstellar discs of these stars to the orbital plane.

  Finally, we make the argument that RW~Aur\,AB is a physical binary.

        %%%%%%%%%%%%%%%%%%%%%%%%%%%%%%%%%%%%%%%%%%%%%%%%%%%%%%%%%%%%%%%%%

\section*{Acknowledgements}

We thank I.~Antokhin, M.~Burlak, D.~Cheryasov, A.~Gusev, K.~Malanchev, O.~Vozyakova as well as the staff of the Caucasian Mountain Observatory headed by N.~Shatsky for the help with observations. We also thank the anonymous referee for benevolent and useful remarks. This research has made use of the SIMBAD database, operated at CDS, Strasbourg, France as well as the VALD \citep{VALD-15} and NIST \citep{NIST-ASD} databases. The study of AD (observations, data reduction, interpretation) and SL (interpretation) and BS (polarimetric observations and data reduction) was conducted under the financial support of the Russian Science Foundation 17-12-01241. Scientific equipment used in this study were bought partially for the funds of the M.~V.~Lomonosov Moscow State University Program of Development.

\section*{Data availability}

The photometric and polarimetric data used in this article are available in online supplementary material.
Other data used in this article will be shared on reasonable request to the corresponding author.
			%%%%%%%%%%%%%%%%%%%%%%%%%%%%%%

\bibliographystyle{mnras}
\bibliography{lamzin2018}
				%%%%%%%%%%%%%%%%%%%%%%%%%

\bsp	% typesetting comment

 \label{lastpage}
\end{document}